\documentclass[journal]{IEEEtran}
\ifCLASSINFOpdf
\else
\fi
\usepackage{amsmath}
\usepackage{algorithm}
\usepackage{mathrsfs}
\usepackage{algpseudocode}
\usepackage{graphicx}
\makeatletter
\def\BState{\State\hskip-\ALG@thistlm}
\makeatother
\begin{document}
%
\title{Efficient Spatial Variation Characterization via Matrix Completion}
%
%
%

\author{Hongge~Chen,
        Duane~Boning,~\IEEEmembership{Fellow,~IEEE,}
        and~Zheng~Zhang
\thanks{H. Chen, D. Boning and Z. Zhang are with Microsystems Technology Laboratories, Massachusetts Institute of Technology, Cambridge,
MA, 02139 USA (e-mail: chenhg@mit.edu, boning@mtl.mit.edu, z\_zhang@mit.edu).}
}

\maketitle

\begin{abstract}
In this paper, we propose a novel method to estimate and characterize spatial variations on dies or wafers. This new technique exploits recent developments in matrix completion, enabling estimation of spatial variation across wafers or dies with a small number of randomly picked sampling points while still achieving fairly high accuracy. This new approach can be easily generalized, including for estimation of mixed spatial and structure or device type information.
\end{abstract}


%
\IEEEpeerreviewmaketitle

\section{Introduction}
%
%
%
%
\IEEEPARstart{A}{dvanced} semiconductor manufacturing is subject to multiple sources of uncertainty. Integrated circuit fabrication at the nanoscale is not entirely deterministic, and variations in manufacturing lead to significant uncertainty in the behavior of individual electronic devices and of circuits as a whole. In order to predict and reduce the impact of variations in fabrication, various silicon integrated circuit analysis and optimization approaches have been proposed, and they are playing important roles in the semiconductor industry. To accurately model circuit performance and yield, these methods rely on testing data from fabricated devices and circuits. However, integrated circuit testing is not free. Large numbers of test structures, such as ring oscillators, must be carefully designed, fabricated, and measured and thus the cost is considerable. To reduce the cost of measurement, the idea of virtual metrology and related methods have been proposed [1]. With virtual metrology, instead of actually measuring some "expensive" variables, mathematical or statistical models are constructed to predict them from fabrication parameters or sensor data. Here, we focus on the spatial variations.

In semiconductor manufacturing, spatial variation mainly refers to the variations among the devices at the different positions of the same die or wafer. In this paper, we seek to exploit recent developments of another important area, compressed sensing [5][6], to build a virtual metrology system for spatial variation. Namely, we propose to randomly sample some points and predict the larger picture of spatial variation. In compressed sensing, images or signals can be reconstructed by numbers of samples much less than the threshold claimed in the classical Nyquist Theorem. As it is straight forward to regard variation on wafers or dies as a 2D signal, the idea to exploit compressed sensing approaches for modeling spatial variation is promising. Related problems have been proposed and solved in recent literature. A method based on discrete cosine transform and maximum a posteriori (MAP) estimation is studied in [1]. This method obtains the frequency domain coefficients from samples using maximum a posteriori (MAP) estimation and then performs an inverse discrete cosine transform. However, if the matrix to be recovered is $m\times n$, the number of coefficients in discrete cosine transform is also $m\times n$ and we may encounter an $m\times n$ dimensional optimization problem, which is computationally expensive when $m$ and $n$ are large. In our following numerical experiment, for example, $m=256$, $n=144$ and $m\times n = 36864$.

In this paper, we use a new method based on low rank matrix completion, which is capable to handle relatively large scale problems. Additionally, this new method can be easily generalized when more information about the variation is available.

 
\section{Mathematical Formulation}
Instead of using the discrete cosine transform, another compressed sensing approach to recover a 2D signal is matrix completion (since we can treat the variation on a wafer or die as a matrix). More specifically, if we suppose the true matrix is $M$, then given the observations of some of its entries $M_{i,j}$, $(i, j) \in \Omega$ we estimate the whole matrix by finding the matrix with lowest rank under the constraint

\begin{equation}
\begin{gathered}
\min_X \text{rank}(X) \\
 \text{Subject to} \ X_{i,j}=M_{i,j}, (i,j)\in\Omega
\end{gathered}
\end{equation}
Though this might seem to be straight-forward, this problem is known to be NP-hard. To make the problem tractable, an approximation is to focus on the convex relaxation of rank$(X)$, the nuclear norm. The nuclear norm of matrix $X$ , $||X||_*$ , is defined as the sum of the singular values of $X$. The convex relaxed version of the problem is thus
\begin{equation}
\begin{gathered}
\min_X ||X||_* \\
  \text{Subject to} \ X_{i,j}=M_{i,j}, (i,j)\in\Omega
\end{gathered}
\end{equation}
Cand\`es and Recht [2] prove that when $M$ is under certain
conditions, if over $Cn^{5/4}r\text{log}n$ randomly picked entries of $M$ are observed, the above problem's solution $X^*$ will be equal to $M$, with very high probability. Here we assume $n\geq m $and the rank of $M$ is $r$ . If we can assume that there exists noise in the measurement, this problem can be further relaxed to a Lagrangian form
\begin{equation}
\min_X ||X||_*+\frac{\lambda}{2}\sum_{(i,j)\in\Omega}|X_{i,j}-M_{i,j}|^2
\end{equation}
with $\lambda\rightarrow \infty$.

Ma, Goldfarb and Chen [3] proposed a fast fixed point continuation algorithm to solve (4), a more general version of (3), based on singular value thresholding.
\begin{equation}
\min_X \mu||X||_*+\frac{1}{2}||\mathscr{A}(X)-b||_2^2,\ \mu\rightarrow 0.
\end{equation}
Here $\mathscr{A}(X)$ is a linear transform mapping $X$ to a vector. It is easy to see that (3) is a special case of $(4)$ with $\mu=1/\lambda$.
\begin{algorithm}
\caption{Solving the minimization in (4)}\label{euclid}
\begin{algorithmic}[1]

\State $\text{Initialized} X=X_0, \mu_1>\mu_2> ... >\mu_{final}$
\For{$\mu=\mu_1, \mu_2, ..., \mu_{final}$}
\While{not converge}
	\State Pick $\tau>0$
	\State $Y=X-\tau\mathscr{A}^*(\mathscr{A}(X)-b)$
	\State Do SVD: $Y=U\text{diag}\{\sigma_i\}V^T$
	\State For each $\sigma_i$, $s_{\tau\mu}(\sigma_i)=\text{max}\{\sigma_i-\tau\mu, 0\}$
	\State $X=U\text{diag}\{s_{\tau\mu}(\sigma_i)\}V^T$ 
\EndWhile
\EndFor
\end{algorithmic}
\end{algorithm}
\section{Numerical Experiments}
In this section, we present numerical experiments of this matrix completion method on real silicon measurement data. Our data are contact resistance measurements from 24 chips fabricated in a 90 nm CMOS process [4]. We have $256\times144 = 36864$ measurements on each chip.
\begin{figure}[h]
\includegraphics[width=9cm]{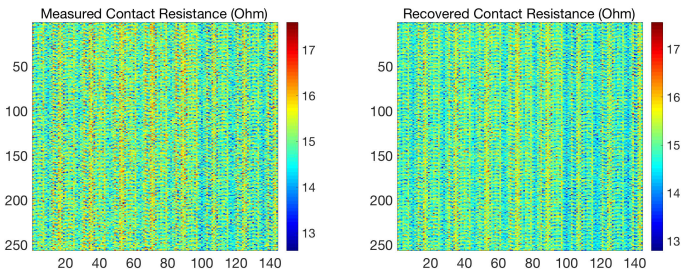}
\caption{Measured contact resistance and recovered contact resistance using matrix completion with 40\% (14746 out of 36864) entries observed.}
\end{figure}
Fig. 1 shows the measured contact resistance matrix $M$, and the recovered contact resistance $X$. Here we obtain $X$ by low-rank matrix completion with 40\% of the entries selected uniformly at random from $M$, and $\mu_{final}= 0.001$. The relative error is $||X-M||_F/||M||_F=0.48\%$, where the Frobenius norm of matrix $X$ is defined as $||X||_F=(\sum_{i,j}X_{i,j}^2)^{1/2}.$ As expected, the relative error decreases as we sample more entries on the matrix.

Additionally, with the algorithm solving (4), we can generalize (3) and add regularization terms when more information is available. For example, in our data we have different layout patterns or device types, and we can add a small regularization term to (3) to exploit such additional information. Among the un-observed entries, we can make approximate estimates if there exist entries of the same layout pattern or device type that are observed. We denote these entries $\Omega'$. The approximate estimation for the device type component (complementing the spatial component) can be determined by the average of the observed entries with the same type. The resulting minimization problem is shown in (5) and we can easily transform it into (4) 
\begin{equation}
\begin{gathered}
\min_X ||X||_*+\frac{\lambda}{2}\sum_{(i, j)\in\Omega}|X_{i,j}-M_{i, j}|^2 \\
 +\frac{\eta}{2}\sum_{(i', j')\in\Omega'}|X_{i', j'}-\hat{X}_{type(i',j')}|^2
\end{gathered}
\end{equation}
Here $\hat{X}_{type(i', j')}$ is the average of the observed entries with the same type with $(i', j')$.

In Figure 2, we fix $\lambda$ at 100, which means $\mu_{final}=0.01$, and sample rate at 40\%, while changing $\eta$. It is seen that with information about layout pattern types and with  $\eta$ properly chosen, we can further reduce the error.
 \begin{figure}[h]
\includegraphics[width=9cm]{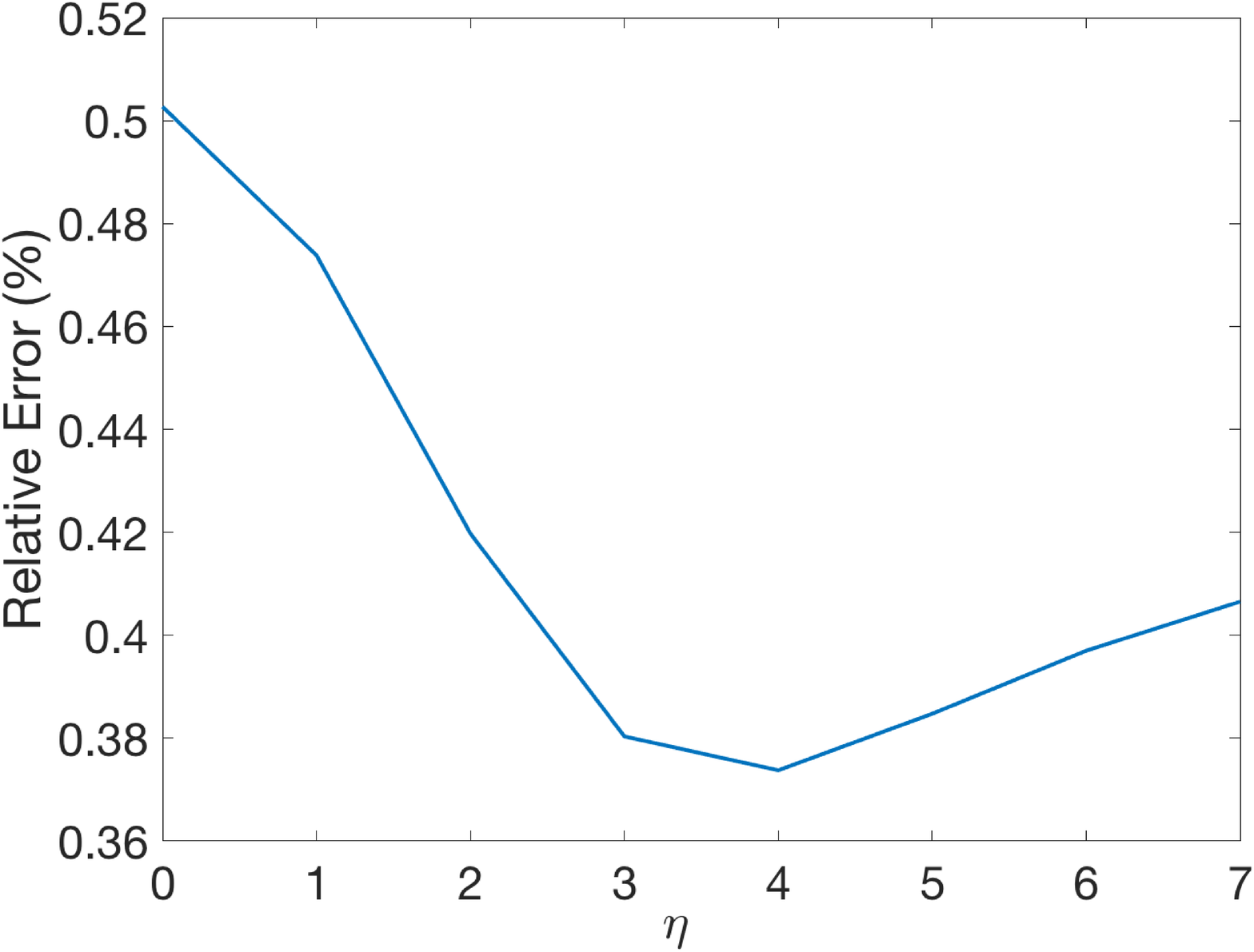}
\caption{Relative error obtained by different $\eta$. Here $\lambda$ is fixed at 100 ($\mu_{final}=0.01$) and sample rate at 40\%.}
\end{figure}
The use of matrix completion opens the door to new approaches for virtual sensing of spatial and device type variation. Future work will further explore and compare efficiency of our results with existing methods, seek to understand limitations and improve the new approach.
\end{document}